\documentclass[english,pra,twocolumn, showpacs,superscriptaddress]{revtex4}
\usepackage[T1]{fontenc}
\usepackage[latin9]{inputenc}
\usepackage{amstext}
\usepackage{amssymb}
\usepackage{esint}
\usepackage{graphicx}
\usepackage[colorlinks,linkcolor=blue,citecolor=blue]{hyperref}
\usepackage{babel}

\makeatletter
\@ifundefined{textcolor}{}
{
 \definecolor{BLACK}{gray}{0}
 \definecolor{WHITE}{gray}{1}
 \definecolor{RED}{rgb}{1,0,0}
 \definecolor{GREEN}{rgb}{0,1,0}
 \definecolor{BLUE}{rgb}{0,0,1}
 \definecolor{CYAN}{cmyk}{1,0,0,0}
 \definecolor{MAGENTA}{cmyk}{0,1,0,0}
 \definecolor{YELLOW}{cmyk}{0,0,1,0}
 }
\makeatother

\begin{document}

\title{Photon-assisted Landau-Zener transition: Role of coherent
superposition states}
\author{Zhe Sun}
\email{sunzhe@hznu.edu.cn}
\affiliation{Advanced Science Institute, RIKEN, Wako-shi, Saitama 351-0198, Japan}
\affiliation{Department of Physics, Hangzhou Normal University, Hangzhou 310036, China}
\author{Jian Ma}
\affiliation{Advanced Science Institute, RIKEN, Wako-shi, Saitama 351-0198, Japan}
\affiliation{Zhejiang Institute of Modern Physics, Department of Physics, Zhejiang
University, Hangzhou 310027, China}
\author{Xiaoguang Wang}
\affiliation{Advanced Science Institute, RIKEN, Wako-shi, Saitama 351-0198, Japan}
\affiliation{Zhejiang Institute of Modern Physics, Department of Physics, Zhejiang
University, Hangzhou 310027, China}
\author{Franco Nori}
\affiliation{Advanced Science Institute, RIKEN, Wako-shi, Saitama 351-0198, Japan}
\affiliation{Physics Department, The University of Michigan, Ann Arbor, Michigan
48109-1040, USA}

\begin{abstract}
We investigate a Landau-Zener (LZ) transition process modeled by a quantum
two-level system (TLS) coupled to a photon mode when the bias energy is
varied linearly in time. The initial state of the photon field is assumed to
be a superposition of coherent states, leading to a more intricate LZ
transition. Applying the rotating-wave approximation (RWA), analytical
results are obtained revealing the enhancement of the LZ probability by
increasing the average photon number. We also consider the creation of
entanglement and the change of photon statistics during the LZ process.
Without the RWA, we find some qualitative differences of the LZ dynamics
from the RWA results, e.g., the average photon number no longer
monotonically enhances the LZ probability. The ramifications and
implications of these results are explored.
\end{abstract}

\maketitle



\section{Introduction}

Landau-Zener (LZ) transitions involve a quantum two-level system (TLS) with
a constant coupling strength $\Delta $ between two adiabatic energy levels.
A control parameter is swept at a constant velocity $v$, so that an avoided
crossing of energy levels occurs, and provides the probability that the
system will stay in an adiabatic state. LZ transitions have attracted
considerable attention theoretically (see, e.g., Refs.\thinspace \thinspace
\cite%
{LZ,Theoretical-1,Theoretical-2,Theoretical-3,Theoretical-4,Theoretical-5})
and experimentally\thinspace (see, e.g., Refs.\thinspace \cite%
{experiment1,experiment2,experiment3}).

In a variety of physical areas, LZ processes play an important role, e.g.,
in artificial atoms~\cite{atoms} and Bose-Einstein condensates in optical
lattices~\cite{BEC}. Especially in superconducting circuits which can behave
like controllable quantum TLSs~\cite{super1,super2,super3,super4}, LZ and
Landau-Zener-St\"{u}ckelberg (LZS) problems have been studied by several
groups\thinspace \cite{LZ,super5,super6,super7,super8,super9,super10}. The
standard LZ problem for an isolated TLS can be solved exactly. For some
many-level systems, the LZ transition probability can also be calculated
exactly for some initial states\thinspace \cite{Shytov,Saito,Altland}. LZ
transitions controlled by classical fields are considered in Ref.\thinspace
\cite{Wubs}, and in a quantum photon field the authors of Ref.\thinspace
\cite{Keeling} found that varying the LZ sweep rate produces collapses and
revivals of the coherent field amplitude.

Naturally, a quantum TLS is influenced by its environment, and therefore
there have been many studies about the dissipative LZ problem. Exact results
are available at zero temperature\thinspace \cite%
{Wubs-Saito-zeroT,Ind-crossing1}, and various numerical methods have been
employed to study the cases at finite temperatures\thinspace \cite%
{Numerical-1,Numerical-2,Numerical-3}. The nonmonotonic dependence of the
LZ probability on the sweep velocity was studied in Ref.\thinspace \cite%
{Numerical-1} using numerical methods. Environment parameters, such as
temperature, can exponentially enhance the coherent oscillations generated
at a LZ transition\thinspace \cite{Numerical-3}.

It is interesting to replace the classical coupling $\Delta $ by a fully
quantum-field coupling; then the TLS and the field form a whole composite
quantum system. In this paper we consider a quantum TLS coupled to a photon
mode when the bias energy is varied linearly in time.

\setlength \parskip{0pt} Coherent superpositions of coherent states, like $%
\left\vert \psi \left( 0\right) \right\rangle _{\text{ph}}=\left( \left\vert
\alpha \right\rangle +e^{i\theta }\left\vert -\alpha \right\rangle \right)
/N_{\theta }$, with the normalization constant $N_{\theta }^{2}=2(1+\cos
\theta e^{-2|\alpha |^{2}})$, have attracted extensive interest as a
distinct class of nonclassical states with interesting properties. For a
large amplitude $\alpha $, these can be interpreted as quantum
superpositions of two macroscopically-distinguishable states, the so-called
Schr\"{o}dinger cat states. Such states can be prepared in various systems
and play an important role in fundamental tests of quantum theory and in
many quantum-information-processing tasks~\cite%
{cat-state1,cat-state2,cat-state3}, including quantum computation\thinspace
\cite{computation-cs}, quantum teleportation\thinspace \cite{telep-cs}, and
precision measurements\thinspace \cite{noon-cs,metrology-cs}.

We aim to discover the effect of the initial superposition of coherent
states on the LZ transition. The increasing average photon number may
enhance the LZ probability. We also focus on the effect of the LZ process on
the quantum properties of the whole system, including entanglement creation
and changing the photon distribution.

By applying a rotating-wave approximation (RWA), we obtain analytical
results which reveal the enhancement of the LZ probability when increasing
the average photon number. Whereas, without the RWA we find some qualitative
differences of the LZ dynamics from the RWA results; e.g., there are two
stages of the LZ transition and the final LZ probability no longer
monotonically depends on the average photon number.

This paper is organized as follows. In Sec.\thinspace II, we introduce the
standard LZ model and the quantized LZ model considered in this paper. In
Sec.\thinspace III, by employing the RWA, we analytically calculate the LZ
probability, the entanglement between the TLS and the field, and the photon
statistics characterized by the Mandel parameter $Q$. Numerical results are
shown in this section to confirm the analytical results. In Sec.\thinspace
IV, \textit{without }the RWA, numerical and analytical results are given to
compare with the RWA results. The thermal state of the photon field is also
considered, in order to compare with the case of a superposition of coherent
states in which the photon distribution is Poissonian. Finally, we present
the conclusions.

\section{ Hamiltonian of the Landau-Zener transition}

Let us first briefly gather together several results that will be used in
this work. The standard LZ problem, for an isolated quantum TLS driven by a
\textit{classical} field, is described by the Hamiltonian
\begin{equation}
H=-\frac{vt}{2}\sigma _{z}-\frac{\Delta }{2}\sigma _{x},
\label{Hlz_standard}
\end{equation}%
in terms of the Pauli matrices $\sigma _{x,z}$, and $\sigma _{x}=\sigma
_{+}+\sigma _{-}$ ($\hbar =1$ is assumed throughout). Let the states $%
\left\vert \uparrow \right\rangle $ and $\left\vert \downarrow \right\rangle
$ denote the eigenstates of $\sigma _{z}$, i.e., the\ so-called diabatic
states with energies $\pm vt/2$ which cross at $t=0$. The constant $v$ is
the sweep velocity, by which the energies of the diabatic states cross. The
coupling $\Delta $ denotes the interaction between the two diabatic states,
which are chosen to be positive and time independent. For $\Delta \neq 0$,
the diabatic states are not eigenstates of the Hamiltonian in Eq.\thinspace (%
\ref{Hlz_standard}), and the avoided-level crossing appears between the
adiabatic energies $E_{\pm }\left( t\right) =\pm \left[ \left( vt\right)
^{2}+\Delta ^{2}\right] /2$ at $t=0$. Thus, generally, a population transfer
is induced. Asymptotically, for times $|t|\gg \Delta /v$, the diabatic
states coincide with the adiabatic states. The LZ problem asks for the
probability of the TLS ending up in the initially unoccupied level, and is
given by $P_{0,\text{LZ}}=1-\exp (-\pi \Delta ^{2}/2v)$, which is an exact
result for all $\Delta $ and $v$. In the adiabatic limit $\Delta ^{2}/v\gg 1$%
, i.e., when the sweep occurs slowly enough, $P_{0,\text{LZ}}$ will saturate
at $1$, which implies that the transfer of population between the adiabatic
eigenstates is prevented by the splitting $\Delta $.

In this work we consider a \textit{quantized} LZ Hamiltonian describing the
coupling of a quantum TLS to a single-photon-field mode. The Hamiltonian
reads
\begin{equation}
H=\frac{\omega _{0}}{2}\sigma _{z}+\omega a^{\dagger }a-\frac{vt}{2}\sigma
_{z}-\frac{\Delta }{2}\sigma _{x}\left( a+a^{\dagger }\right) ,
\label{H-nonRWA}
\end{equation}%
where the operator $a$ ($a^{\dagger }$) annihilates (creates) a photon in
the field mode with frequency $\omega $, and the energy bias of the TLS is
denoted by $\omega _{0}$. In this paper, the resonance case $\omega
_{0}=\omega $ is considered. Hence we rewrite the total Hamiltonian in a
rotating frame defined by the operator $\hat{N}=a^{\dagger }a+\sigma _{z}/2$%
, at the frequency $\omega $.

In the weak-coupling regime, where the coupling is at least an order of
magnitude less than the energy frequency, i.e., $\Delta <0.1\omega $, one
can employ a RWA, and the Hamiltonian becomes
\begin{equation}
H=-\frac{vt}{2}\sigma _{z}-\frac{\Delta }{2}\left( a\sigma _{+}+a^{\dagger
}\sigma _{-}\right) .  \label{H_LZQ}
\end{equation}%
Now the system is modeled in terms of a time-dependent Jaynes-Cummings
Hamiltonian\thinspace \cite{Larson1,Larson2,Keeling}, which can also be
called the Landau-Zener-Jaynes-Cummings (LZ-JC) model. Note that the
operator $\hat{N}=a^{\dagger }a+\sigma _{z}/2$ is conserved by this
Hamiltonian.

When considering a finite detuning $\delta \omega =\omega _{0}-\omega $, the
RWA is usually justified in the condition that $\left\vert \delta \omega
\right\vert \ll \omega _{0}+\omega $. Then a constant energy bias $\delta
\omega $ is added to the adiabatic states, and the LZ-JC Hamiltonian becomes
$H=-\frac{vt-\delta \omega }{2}\sigma _{z}-\frac{\Delta }{2}\left( a\sigma
_{+}+a^{\dagger }\sigma _{-}\right) $. The finite detuning $\delta \omega $
gives no change to the LZ process except translating the time when the LZ
transition occurs, from $t=0$ to $t=\delta \omega /v$.

This kind of model was considered in Ref.\thinspace \cite{Larson1}, where
highly nonclassical sub-Poissonian states were found. Some later
work\thinspace \cite{Larson2,Saito} mentioned the LZ problem in the JC and
Rabi models. A superconducting qubit coupled to a transmission-line
resonator\thinspace \cite{c-QED1} can be described by the quantized LZ
Hamiltonian in Eqs.\thinspace (\ref{H-nonRWA}) and (\ref{H_LZQ}) when the
transition frequency of the charge (flux) qubit is varied linearly in time
by a driving charge (magnetic flux). Furthermore, in this kind of circuit
quantum electrodynamics, the strong coupling can be obtained, which allows
the cases beyond the RWA\thinspace \cite{c-QED2}.

\section{Landau-Zener transition processes, entanglement creation and photon
distribution with the RWA}

Let us assume that the initial state of the total system is in a direct
product form, and the TLS initially starts from $|\uparrow \rangle $ and
that the photon field starts from a Fock state, then the initial state of
the whole system $\left\vert \psi \left( 0\right) \right\rangle _{\text{tot}}
$ $=\left\vert \uparrow \right\rangle \otimes \left\vert n\right\rangle $.
Under the LZ-JC Hamiltonian in Eq.~(\ref{H_LZQ}), the state at time $t$
becomes
\begin{equation}
\left\vert \psi \left( t\right) \right\rangle _{\text{tot}}=A_{n}\left(
t\right) \left\vert \uparrow n\right\rangle +B_{n}\left( t\right) \left\vert
\downarrow n+1\right\rangle ,
\end{equation}%
where, the time-dependent coefficients $A_{n}\left( t\right) $ and $%
B_{n}\left( t\right) $ are the solutions of a second-order Weber equation
and in the form of combinations of parabolic cylinder functions \cite{LZ}.
Explicitly, the coefficients
\begin{eqnarray}
A_{n}(t) &=&\sum_{\pm }\mu _{n,\pm }D_{-1-i\delta _{n}}\left( \pm
Z_{t}\right) ,  \nonumber \\
B_{n}(t) &=&\sum_{\pm }\nu _{n,\pm }D_{-i\delta _{n}}\left( \pm Z_{t}\right)
,
\end{eqnarray}%
where $D_{-1-i\delta _{n}}\left( \pm Z_{t}\right) $ [$D_{-i\delta
_{n}}\left( \pm Z_{t}\right) $] are the parabolic cylinder functions and the
parameters $Z_{t}=-\sqrt{2}e^{i\pi /4}\sqrt{v/2}t$ and $\delta _{n}=\Delta
^{2}\left( n+1\right) /(4v)$, because when the field mode is occupied by $n$
photons, the splitting $\Delta $ is enhanced by a factor $\sqrt{n+1}$, as
compared with the standard LZ model. The parameters $\nu _{n,\pm }$ and $\mu
_{n,\pm }$ satisfy $\nu _{n,\pm }=\mp \mu _{n,\pm }e^{-i\pi /4}/\sqrt{\delta
_{n}}$. Moreover, if the bias energy $vt_{0}$ is finite, then we have $\mu
_{n,+}=D_{-i\delta _{n}}\left( -Z_{t_{0}}\right) /[D_{-i\delta _{n}}\left(
-Z_{t_{0}}\right) D_{-1-i\delta _{n}}\left( Z_{t_{0}}\right) +D_{-i\delta
_{n}}\left( Z_{t_{0}}\right) D_{-1-i\delta _{n}}\left( -Z_{t_{0}}\right) ]$
and $\mu _{n,-}=\mu _{n,+}D_{-i\delta _{n}}\left( Z_{t_{0}}\right)
/D_{-i\delta _{n}}\left( -Z_{t_{0}}\right) $. Usually, one considers the
limit cases when the bias energy $vt$ is switched from a large negative
value to a large positive value; then with the asymptotes of the parabolic
cylinder functions, we find $A_{n}(t)\approx \exp (-\pi \delta _{n})$, which
implies that for an initial state $\left\vert \uparrow n\right\rangle $, the
final probability of the TLS staying in $\left\vert \uparrow \right\rangle $
is $P_{\uparrow ,n}=P_{\uparrow ,0}\exp (-\pi \Delta ^{2}n/2v)$, with $%
P_{\uparrow ,0}=\exp (-\pi \Delta ^{2}/2v)$ denoting the vacuum case.

When the initial state of the field is a superposition of Fock states, the
initial state of the whole system reads
\begin{equation}
\left\vert \psi \left( 0\right) \right\rangle _{\text{tot}}=\left\vert
\uparrow \right\rangle \otimes \sum_{n=0}^{\infty }C_{n}\left\vert
n\right\rangle ,  \label{st_initial}
\end{equation}%
where the superposition parameter $C_{n}$ satisfies $\sum_{n=0}^{\infty
}\left\vert C_{n}\right\vert ^{2}=1$. Then, multi level LZ transitions are
expected at the avoided-level crossings of the diabatic energy levels, and
we sketch the energy-level diagram in Fig.\thinspace 1\thinspace (a). The
final probability of the TLS staying in $\left\vert \uparrow \right\rangle $
becomes an average of $P_{\uparrow ,n}$, and then the final LZ transition
probability at $t=\infty $ reads
\begin{equation}
P_{\text{LZ}}\left( \infty \right) =1-P_{\uparrow ,0}\sum_{n=0}^{\infty
}\left\vert C_{n}\right\vert ^{2}\exp \left( -\frac{\pi \Delta ^{2}n}{2v}%
\right) .  \label{P_LZ_n}
\end{equation}

\subsection{LZ processes for coherent superposition states}

Hereafter we focus on the nonclassical properties associated with the
following superposition of coherent states:
\begin{equation}
\left\vert \psi \left( 0\right) \right\rangle _{\text{ph}}=\frac{1}{%
N_{\theta }}\left( \left\vert \alpha \right\rangle +e^{i\theta }\left\vert
-\alpha \right\rangle \right) ,  \label{st_ph}
\end{equation}%
where $N_{\theta }^{2}=2(1+\cos \theta e^{-2|\alpha |^{2}})$ is a
normalization constant and, for simplicity, we assume $\alpha $ to be real.
Such states are superpositions of classically distinguishable states and
involve fundamentally nonclassical properties. Therefore, they are important
for investigating fundamental tests of quantum theory and in many quantum
information processing tasks~\cite{cat-state1,cat-state2,cat-state3}.

As shown in Refs.\thinspace \cite{YS-state,YS-creation}, for an anharmonic
oscillator with the Hamiltonian $H=\upsilon \left( a^{\dagger }a+\frac{1}{2}%
\right) +\chi \left( a^{\dagger }a\right) ^{2}$, an initial coherent state $%
\left\vert \alpha \right\rangle $ will evolve into the coherent
superposition states in Eq.\thinspace (\ref{st_ph}) with a superposition
phase $\theta $ corresponding to different evolution times. And at some
special time such as $t=\pi /\left( 2\chi \right) $, the so-called
\textquotedblleft Yurke-Stoler\textquotedblright\ coherent state with $%
\theta =\pi /2$ can be achieved. In an optical system, the coherent
superposition states with large amplitude can be generated by using homodyne
detection and photon number states as resources\thinspace \cite{cat-state3},
and the superposition phase $\theta $ is related to the photon numbers.

Recalling the initial state in Eq.~(\ref{st_initial}), the superposition
coefficient becomes
\begin{equation}
C_{n}=\frac{1}{N_{\theta }}\exp \left( -\frac{\left\vert \alpha \right\vert
^{2}}{2}\right) \frac{\alpha ^{n}[1+(-1)^{n}]}{\sqrt{n!}}.
\end{equation}%
It follows that we can calculate the final LZ transition probability exactly
\begin{equation}
P_{\text{LZ}}\left( \infty \right) =1-\frac{2P_{\uparrow ,0}}{N_{\theta
}^{2}e^{\left\vert \alpha \right\vert ^{2}}}\left( e^{\left\vert \alpha
\right\vert ^{2}P_{\uparrow ,0}}+\cos \theta e^{-\left\vert \alpha
\right\vert ^{2}P_{\uparrow ,0}}\right) .  \label{P_LZ_theta}
\end{equation}%
It turns out that $P_{\text{LZ}}\left( \infty \right) $ now depends on the
initial conditions $\left\vert \alpha \right\vert ^{2}$ and $\theta $; the
former is associated with the average photon of field and the latter
determines the types of superpositions. Obviously, for $\left\vert \alpha
\right\vert ^{2}=0$ and $\cos \theta \neq -1$, then $P_{\text{LZ}}\left(
\infty \right) =1-P_{\uparrow ,0}$, corresponding to the standard LZ
probability. Whereas in the limit $\left\vert \alpha \right\vert
^{2}\rightarrow \infty $, with finite ratio $\Delta ^{2}/v\neq 0$, $P_{\text{%
LZ}}\left( \infty \right) $ will tend to unity monotonically.

\subsubsection{Yurke-Stoler state}

When $\theta =\pi /2,$ the superposition state is the so-called Yurke-Stoler
(YS) coherent state \cite{YS-state}: $\left\vert \alpha \right\rangle _{%
\text{YS}}=\left( \left\vert \alpha \right\rangle +i\left\vert -\alpha
\right\rangle \right) /N_{\pi /2}$. The average photon number of $\left\vert
\alpha \right\rangle _{\text{YS}}$ is $\left\vert \alpha \right\vert ^{2}$.
Thus, the LZ probability for the YS state becomes
\begin{equation}
P_{\text{LZ}}\left( \infty \right) =1-P_{\uparrow ,0}\exp \left[ -\left\vert
\alpha \right\vert ^{2}(1-P_{\uparrow ,0})\right] ,  \label{P_LZ_YS}
\end{equation}%
which reveals the dependence of $P_{\text{LZ}}\left( \infty \right) $ on the
ratio $\Delta ^{2}/v$ and the average photon number $\left\vert \alpha
\right\vert ^{2}$. Obviously, enhancing $\left\vert \alpha \right\vert ^{2}$
and the ratio $\Delta ^{2}/v$ will increase the final LZ probability $P_{%
\text{LZ}}\left( \infty \right) $.

\subsubsection{Even coherent state}

When $\theta =0$, the photon state is the so-called \textquotedblleft even
coherent state\textquotedblright : $\left\vert \alpha \right\rangle
_{+}=\left( \left\vert \alpha \right\rangle +\left\vert -\alpha
\right\rangle \right) /N_{0}$, with $N_{0}^{2}=2(1+e^{-2|\alpha |^{2}})$.
This state refers to the fact that the photon number distribution is nonzero 
only for even photon numbers with the average photon number $\bar{n}%
=2\left\vert \alpha \right\vert ^{2}\left( 1-e^{-2|\alpha |^{2}}\right)
/N_{0}^{2}$. Then the final probability
\begin{equation}
P_{\text{LZ},+}\left( \infty \right) =1-P_{\uparrow ,0}\frac{\cosh
(\left\vert \alpha \right\vert ^{2}P_{\uparrow ,0})}{\cosh \left\vert \alpha
\right\vert ^{2}}.
\end{equation}

\subsubsection{Odd coherent state}

When $\theta =\pi $, the photon state is an \textquotedblleft odd coherent
state\textquotedblright : $\left\vert \alpha \right\rangle _{-}=\frac{1}{%
N_{\pi }}\left( \left\vert \alpha \right\rangle -\left\vert -\alpha
\right\rangle \right) $, with $N_{\pi }^{2}=2(1-e^{-2|\alpha |^{2}})$, for
which only an odd number of photons have a nonzero probability and the
average photon number $\bar{n}=2\left\vert \alpha \right\vert ^{2}\left(
1+e^{-2|\alpha |^{2}}\right) /N_{\pi }^{2}$. Then the LZ probability becomes
\begin{equation}
P_{\text{LZ},-}\left( \infty \right) =1-P_{\uparrow ,0}\frac{\sinh
(\left\vert \alpha \right\vert ^{2}P_{\uparrow ,0})}{\sinh \left\vert \alpha
\right\vert ^{2}}.
\end{equation}%
When $\left\vert \alpha \right\vert ^{2}$ approaches zero, we have $P_{\text{%
LZ},-}\left( \infty \right) \rightarrow 1-P_{\uparrow ,0}^{2}$, because the
odd coherent state $\left\vert \alpha \right\rangle _{-}$ tends to the Fock
state $\left\vert 1\right\rangle $.
\begin{figure}[tbp]
\includegraphics[bb=-17 217 579 622,
width=0.49\textwidth,height=0.42\textwidth]{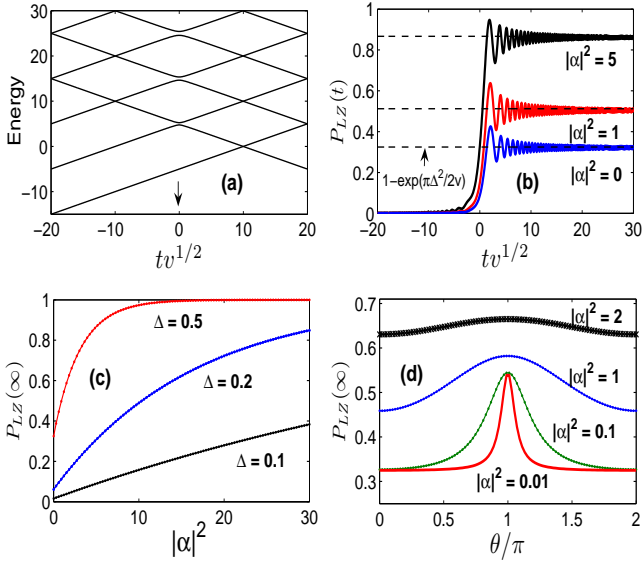}
\caption{(Color online) (a) Adiabatic energy levels of the quantized LZ
Hamiltonian~(\protect\ref{H_LZQ}) within the RWA. The coupling $\Delta =0.5$
and frequency $\protect\omega =10$ are both in units of $\protect\sqrt{v}$,
also $\hbar =1$. The arrow marks the point where the avoided crossings are
located. (b) LZ probability $P_{\text{LZ}}(t)$ as a function of time, in
units of $1/\protect\sqrt{v}$, for the coupling $\Delta =0.5$, and for
various values of the average photon number $|\protect\alpha |^{2}$. The
horizontal black dashed lines show the analytical results in Eq.~(\protect
\ref{P_LZ_YS}). (c) Final LZ transition probability $P_{\text{LZ}}(\infty )$
versus average photon number $|\protect\alpha |^{2}$ for various couplings $%
\Delta =0.1,~0.2$, and $0.5$. (d) Final LZ probability $P_{\text{LZ}}(\infty
)$ as a function of the superposition parameter $\protect\theta $, for the
coupling $\Delta =0.5$ and various values of $|\protect\alpha |^{2}$.}
\end{figure}

\subsubsection{Numerical results of LZ processes for various types of
coherent superposition states}

Let us first focus on the YS coherent state and numerically study the LZ
processes. In Fig.\thinspace 1\thinspace (b), we show that for a weak
coupling $\Delta =0.5$ (in units of $\sqrt{v}$), the LZ transition occurs
near the point $t=0$, at the avoided crossings. The coherent oscillation of $%
P_{\text{LZ}}(t)$ is enhanced by increasing the average photon number $%
\left\vert \alpha \right\vert ^{2}$. Figure\thinspace 1\thinspace (c) shows
the final probability $P_{\text{LZ}}(\infty )$ as a function of $\left\vert
\alpha \right\vert ^{2}$ for different couplings $\Delta $. As expected in
the analytical results, for increasing $\left\vert \alpha \right\vert ^{2}$
we find that $P_{\text{LZ}}(\infty )$ tends to $1$ monotonically, and larger
couplings $\Delta $ accelerate this increase. The superposition parameter $%
\theta $ gives a periodic contribution to $P_{\text{LZ}}(\infty )$, as shown
in Fig.\thinspace 1\thinspace (d). For small average photon numbers $%
\left\vert \alpha \right\vert ^{2}<2$, there is a clear maximum of $P_{\text{%
LZ}}(\infty )$ at $\theta =\pi $ (odd coherent state), which implies that
not only the average photon number but also the proper superpositions of
coherent states can enhance the LZ probabilities. When $\left\vert \alpha
\right\vert ^{2}$ is larger, however, there is hardly any effect of the
superposition parameter $\theta $. Then all the superpositions of coherent
states provide the same asymptotic value of $P_{\text{LZ}}(\infty )$, as
shown in Fig.\thinspace 1\thinspace (d).

\subsection{Entanglement between the quantum TLS and the photon field}

Due to the coupling terms in the LZ-JC Hamiltonian, the dynamics will
produce entanglement between the quantum TLS and the photon field. The aim
of this section is to reveal\ the connection between the LZ transition and
the entanglement creation. The concept of purity can be employed to
characterize entanglement. Based on the reduced density of TLSs, purity is
determined by the linear entropy, defined by $E_{l}\left( t\right) =1-$Tr$%
\rho _{_{\text{TLS}}}^{2}\left( t\right) $. In terms of the elements of the
density matrix, we have $E_{l}\left( t\right) =1-\sum_{i,j=\uparrow
,\downarrow }$ $\left\vert \left\langle i\right\vert \rho _{_{\text{TLS}%
}}\left\vert j\right\rangle \right\vert ^{2}$. In the LZ process we find the
matrix elements of the density matrix as a function of time%
\begin{eqnarray}
\left\vert \left\langle \uparrow \right\vert \rho _{_{\text{TLS}}}\left\vert
\downarrow \right\rangle \right\vert ^{2} &=&\frac{-\sin ^{2}\theta }{%
N_{\theta }^{2}e^{2\left\vert \alpha \right\vert ^{2}}}\left\vert
\sum_{n=0}^{\infty }\frac{(-1)^{n}\left\vert \alpha \right\vert
^{2n+1}A_{n+1}B_{n}^{\ast }}{\sqrt{(n+1)}n!}\right\vert ^{2},  \nonumber \\
\left\vert \left\langle \uparrow \right\vert \rho _{_{\text{TLS}}}\left\vert
\uparrow \right\rangle \right\vert ^{2} &=&\left[ 1-P_{\text{LZ}}\left(
t\right) \right] ^{2},  \nonumber \\
\left\vert \left\langle \downarrow \right\vert \rho _{_{\text{TLS}%
}}\left\vert \downarrow \right\rangle \right\vert ^{2} &=&P_{\text{LZ}%
}^{2}\left( t\right) .  \label{rho_reduced}
\end{eqnarray}%
Due to the exact solutions of $A_{n}\left( t\right) $ and $B_{n}\left(
t\right) $, the linear entropy can be analytically obtained. If we choose
even (or odd) coherent states in Eq.\thinspace (\ref{st_ph}) for $\theta =0$
(or $\pi $), there will be a simple form of the linear entropy
\begin{equation}
E_{l}\left( t\right) =2P_{\text{LZ}}\left( t\right) \left[ 1-P_{\text{LZ}%
}\left( t\right) \right] ,
\end{equation}%
which implies that the TLS and the photon field can achieve full
entanglement, in the sense that, after tracing out the photon states, no
coherence between $\left\vert \uparrow \right\rangle $ and $\left\vert
\downarrow \right\rangle $ is left; i.e., the antidiagonal elements in the
reduced density matrix of Eq.\thinspace (\ref{rho_reduced}) vanish. In this
case, the entanglement is absolutely determined by $P_{\text{LZ}}\left(
t\right) $. At finite times,\ when $P_{\text{LZ}}\left( t\right) $ suddenly
jumps to a nonzero value but is less than 1/2, $E_{l}\left( t\right) $
increases fast to a steady value. However, if $P_{\text{LZ}}\left( t\right) $
is larger than 1/2, $E_{l}\left( t\right) $ decreases. The entanglement will
be less perfect if other initial superposition parameters ($\theta \neq
0,\pi $) are chosen, because there will be nonzero off-diagonal elements in
the reduced density matrix. The entanglement dynamics strongly depends on
the LZ transition probability, and this is confirmed by the numerical
results shown in Fig.\thinspace 2.
\begin{figure}[tbp]
\includegraphics[width=0.52\textwidth,height=0.425\textwidth]{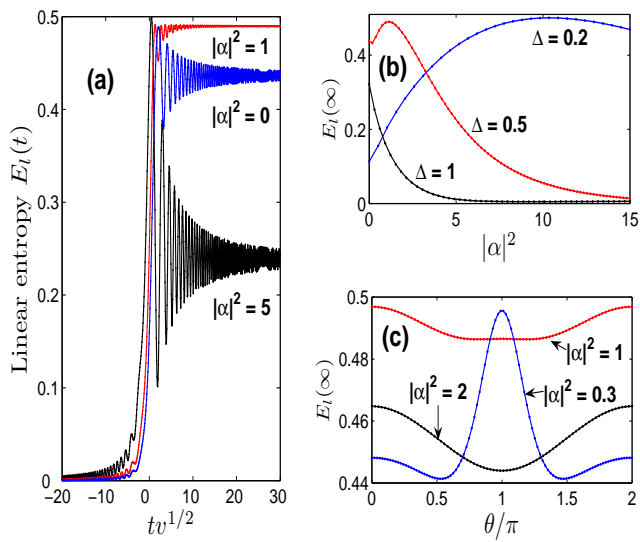}
\caption{(Color online) (a) Linear entropy $E_{l}(t)$ as a function of time
(in units of $1/\protect\sqrt{v}$) for the coupling $\Delta =0.5$ (in units
of $\protect\sqrt{v}$) and for various values of the average photon number $%
|\protect\alpha |^{2}$. (b) Final or long-time linear entropy $E_{l}(\infty )
$ as a function of $|\protect\alpha |^{2}$, for various couplings $\Delta
=0.2,~0.5,$ and $1$. (c) Final linear entropy $E_{l}(\infty )$ versus the
superposition parameter $\protect\theta $, for the coupling $\Delta =0.5$
and various $|\protect\alpha |^{2}$.}
\end{figure}

For an initial YS coherent state, the dynamics of the linear entropy $%
E_{l}\left( t\right) $ for finite times is shown in Fig.\thinspace
2\thinspace (a). The entanglement between the TLS and the photon field is
created when the LZ transition occurs. Analogous to the LZ probability, $%
E_{l}\left( t\right) $ oscillates with time and tends to a steady value.
However, it is not always the case that the larger $\left\vert \alpha
\right\vert ^{2}$ is, the stronger the entanglement will be. In
Fig.\thinspace 2\thinspace (b), we plot the final linear entropy $%
E_{l}(t=\infty )$ versus $\left\vert \alpha \right\vert ^{2}$. Clearly, for
small values of the coupling $\Delta $, the linear entropy $E_{l}(\infty )$
has a maximum. However, for larger couplings, like $\Delta =1$, increasing $%
\left\vert \alpha \right\vert ^{2}$ only suppresses $E_{l}(\infty )$
rapidly. Note that in the large-coupling $\Delta $ limit, $P_{\text{LZ}%
}\left( \infty \right) $ achieves unity adiabatically, which implies that
the final state is separable. In Fig.\thinspace 2\thinspace (c), the linear
entropy $E_{l}(\infty )$ versus the initial superposition parameter $\theta $
for a small coupling $\Delta =0.5$ is considered. The odd coherent state at $%
\theta =\pi $ has a maximal entanglement for a small photon number $%
\left\vert \alpha \right\vert ^{2}=0.3$, whereas it has a minimal
entanglement for a larger photon number $\left\vert \alpha \right\vert ^{2}=2
$.

\subsection{Photon distribution of the field}

One of the best-known nonclassical effects is the generation of
sub-Poissonian (or super-Poissonian) photon statistics of the light
field\thinspace \cite{sub-Poisson,sub-Poisson2}. A coherent state $%
\left\vert \alpha \right\rangle $, which can be regarded as a state with the
\textquotedblleft most\textquotedblright\ classical behavior, yields a
Poissonian distribution, i.e., the variance of the number operator $\hat{n}%
=a^{\dagger }a$ is equal to the mean photon number: $\left( \Delta \hat{n}%
\right) ^{2}=\bar{n}=\left\vert \alpha \right\vert ^{2}$. Mandel\thinspace
\thinspace introduced the $Q$ parameter \cite{sub-Poisson},
\begin{equation}
Q=\frac{\left( \Delta n\right) ^{2}}{\bar{n}}-1,  \label{Q}
\end{equation}%
which characterizes the departure from the Poissonian distribution, i.e.,
the nonclassical property. When $Q=0$, the state is called Poissonian,
while, for $Q>0$ the state is super-Poissonian. If $-1\leq Q<0$, the
statistics is sub-Poissonian. It is known that the Yurke-Stoler coherent
state is Poissonian. However, under time evolution, the parameter $Q$
changes with time as $Q(t)=\left( \Delta \hat{n}\right) _{t}^{2}/\bar{n}%
_{t}-1$. Thus the initial Poissonian may turn into sub-Poissonian or
super-Poissonian. It is interesting to consider how\ the photon distribution
changes during the LZ process.

For the superposition of coherent states in Eq.\thinspace (\ref{st_ph}), at
finite times, we obtain the average photon number
\begin{equation}
\bar{n}_{t}=\bar{n}_{0}+P_{\text{LZ}}\left( t\right) ,
\end{equation}%
where $\bar{n}_{0}=2\left\vert \alpha \right\vert ^{2}\left(
1-e^{-2\left\vert \alpha \right\vert ^{2}}\cos \theta \right) /N_{\theta
}^{2}$ indicates the initial average photon number. The expression given
above reflects that the dynamics of $\bar{n}_{t}$ is dominated by the LZ
transition probability $P_{\text{LZ}}\left( t\right) $. The asymptotic
behavior at infinite time will be $\bar{n}_{\infty }=\bar{n}_{0}+P_{\text{LZ}%
}\left( \infty \right) $. We also obtain the average value of $\hat{n}^{2}$
at $t=\infty $,
\begin{eqnarray}
\left\langle \hat{n}^{2}\right\rangle _{\infty } &=&\frac{-4\left\vert
\alpha \right\vert ^{2}P_{\uparrow ,0}^{2}}{N_{\theta }^{2}e^{\left\vert
\alpha \right\vert ^{2}}}\left( e^{\left\vert \alpha \right\vert
^{2}P_{\uparrow ,0}}-e^{-\left\vert \alpha \right\vert ^{2}P_{\uparrow
,0}}\cos \theta \right)   \nonumber \\
&&+\left\vert \alpha \right\vert ^{4}+3\bar{n}_{0}+P_{\text{LZ}}\left(
\infty \right) ,
\end{eqnarray}%
where $P_{\text{LZ}}\left( \infty \right) $ is shown in Eq.\thinspace (\ref%
{P_LZ_theta}). By the definition in Eq.\thinspace (\ref{Q}), the asymptotic
value of $Q(t)$ becomes $Q(\infty )=\left( \Delta \hat{n}\right) _{\infty
}^{2}/\bar{n}_{\infty }-1$, where the variance $\left( \Delta \hat{n}\right)
_{\infty }^{2}=\left\langle \hat{n}^{2}\right\rangle _{\infty }-\bar{n}%
_{\infty }^{2}$. We shall now concentrate on some limiting cases: (i) when $%
\left\vert \alpha \right\vert ^{2}=0$, for finite LZ rate $\Delta ^{2}/v\neq
0$, \ we have $Q(\infty )<0$, i.e., the LZ transition induces sub-Poissonian
statistics in the photon field; and (ii) when $\left\vert \alpha \right\vert
^{2}$ is very large, we have $Q(\infty )\rightarrow 0$, in which case the LZ
transition has no effect on the photon distribution.
\begin{figure}[tbp]
\includegraphics[width=0.52\textwidth,height=0.425\textwidth]{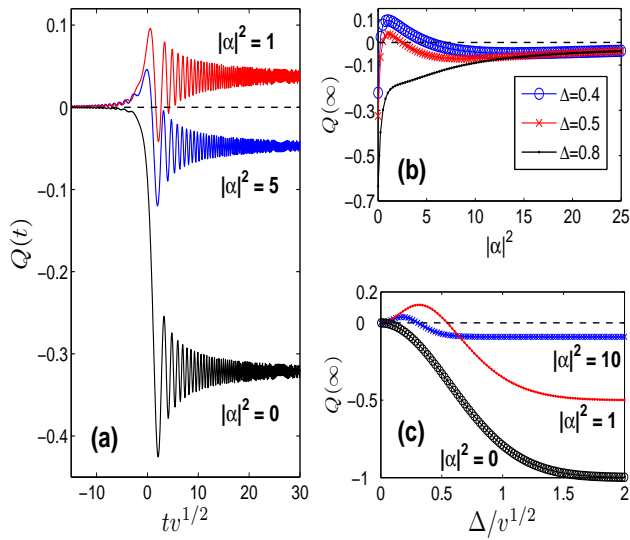}
\caption{(Color online) (a) Mandel parameter $Q(t)$ as a function of time in
units of $1/\protect\sqrt{v}$ for various values of the average photon
number $|\protect\alpha |^{2}$ and the coupling $\Delta =0.5$ in units of $%
\protect\sqrt{v}$. (b) Final Mandel parameter $Q(\infty )$ as a function of $%
|\protect\alpha |^{2}$ for various couplings $\Delta =0.2,~0.5$, and $1$.
(c) Final $Q(\infty )$ versus coupling $\Delta $ for various $|\protect%
\alpha |^{2}$. The horizontal black dashed lines mark $Q=0$ and correspond to a
Poissonian distribution.}
\end{figure}

In Fig.\thinspace 3\thinspace (a), we numerically plot the Mandel parameter $%
Q(t)$ versus time $t$. Obviously, the photon statistics of the field changes
suddenly when the LZ transition occurs. Near the avoided-level-crossing
point, for $\left\vert \alpha \right\vert ^{2}>0$, super-Poissonian and
sub-Poissonian statistics appear alternately. The final Mandel parameter $%
Q(\infty )$ versus $\left\vert \alpha \right\vert ^{2}$ is shown in
Fig.\thinspace 3\thinspace (b), for weak couplings $\Delta $, both
sub-Poissonian and super-Poissonian statistics can appear with increasing $%
\left\vert \alpha \right\vert ^{2}$. For the large $\left\vert \alpha
\right\vert ^{2}$ limit, the photon distribution finally tends to
Poissonian, which is not shown here. We also show $Q(\infty )$ versus the LZ
parameter $\Delta /\sqrt{v}$ in Fig.\thinspace 3\thinspace (c). When $%
\left\vert \alpha \right\vert ^{2}=0$, $Q$ monotonically decays with $\Delta
/\sqrt{v}$ from $0$ to $-1$, whereas, for a finite average photon number
such as $\left\vert \alpha \right\vert ^{2}=1$, super-Poissonian statistics
also appears. However, a large $\left\vert \alpha \right\vert ^{2}$ will
erase the nonclassical effects revealed by the sub-Poissonian (or
super-Poissonian).

\section{Landau-Zener transition without the RWA}

\begin{figure}[tbp]
\includegraphics[bb=1 233 594 614, width=0.49\textwidth, clip]{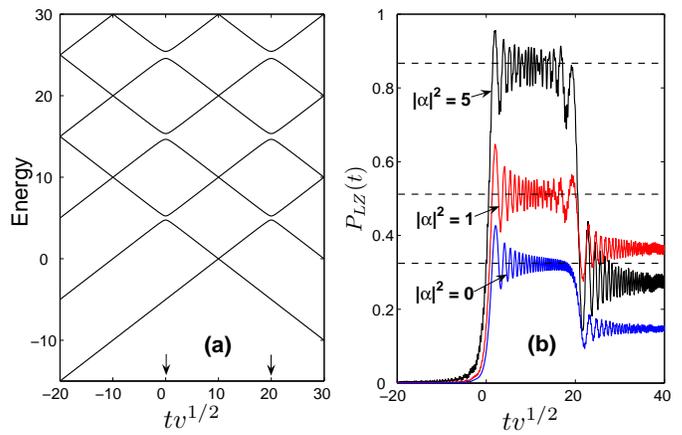}
\caption{(Color online) (a) Adiabatic energy levels of the quantized LZ
Hamiltonian~(\protect\ref{H-nonRWA}) without RWA. The coupling $\Delta =0.5$%
, frequency $\protect\omega =10$, and the parameters are in units of $%
\protect\sqrt{v}$. The vertical arrows show the points where the avoided
crossings are located. (b) LZ probability $P_{\text{LZ}}(t)$ as a function
of time (in units of $1/\protect\sqrt{v}$) for various values of the average
photon number $|\protect\alpha |^{2}$, and for the coupling $\Delta =0.5$
and frequency $\protect\omega =10$. The horizontal black dashed lines show
the analytical results of Eq.~(\protect\ref{P_LZ_YS}) with the RWA which, in
the middle, approximately agree with the first-stage LZ transitions.}
\end{figure}
\begin{figure}[tbp]
\includegraphics[width=0.51\textwidth,height=0.415\textwidth]{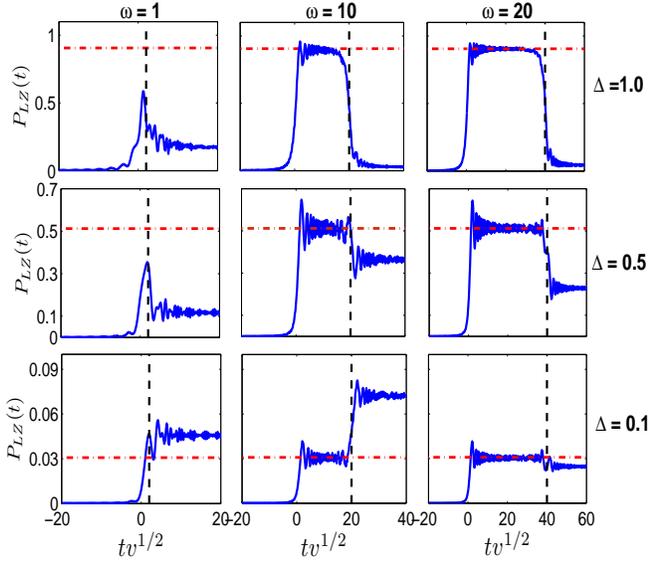}
\caption{(Color online) LZ transition probability $P_{\text{LZ}}(t)$ as a
function of time (in units of $1/\protect\sqrt{v}$) and various values of
the coupling $\Delta $ (in units of $\protect\sqrt{v}$) and frequency $%
\protect\omega $ (in units of $\protect\sqrt{v}$). The average photon number
$|\protect\alpha |^{2}=1$. The horizontal red dash-dot lines show the
analytical results of Eq.~(\protect\ref{P_LZ_YS}) within the RWA. The
vertical black dashed lines indicate the times between the successive
avoided-level crossings, i.e., the two stages of the LZ transitions.}
\end{figure}
In this section, we consider the LZ transition \textit{without }the RWA. A
reasonable comparison between the solutions with and without the RWA will
enable us to understand the contribution from the counter-rotating terms.

\subsection{YS coherent state case}

Without the RWA, due to the counter-rotating terms $\sigma _{+}a^{\dagger }$
and $\sigma _{-}a$, the total number operator $\hat{N}=a^{\dagger }a+\sigma
_{z}/2$ is not conserved by the Hamiltonian in Eq.\thinspace (\ref{H-nonRWA}%
), i.e., $[\hat{N},H]\neq 0$. Then the Hamiltonian (\ref{H-nonRWA}) cannot
be exactly diagonalized and has a fundamentally different energy structure
from the Hamiltonian\thinspace (\ref{H_LZQ}) within the RWA, no matter how
small the coupling $\Delta $ is. From the adiabatic energy spectrum in
Fig.\thinspace 4\thinspace (a), clearly, one can find two groups of
avoided-level crossings, which are significantly different from the RWA case
in Fig.\thinspace 1\thinspace (a). The groups of avoided-level crossing are
formed not only between the states $\left\vert \uparrow n\right\rangle $ and
$\left\vert \downarrow n+1\right\rangle $ but also between $\left\vert
\uparrow n\right\rangle $ and $\left\vert \downarrow n-1\right\rangle $
(when $n\geq 2$), with level splittings $\Delta \sqrt{n+1}$ and $\Delta
\sqrt{n}$, respectively. Note that the two groups of avoided crossings are
approximately independent whenever the time between the successive avoided
level crossings, $t_{\text{cross}}=2\omega /v$, exceeds the duration of an
individual LZ transition, $\tau _{\text{LZ}}\sim \max \left\{ 1/\sqrt{v}%
,\Delta /v\right\} $ \cite{LZ-time,Ind-crossing2}. For our multilevel LZ
problem, the couplings become $\Delta _{n}=\Delta \sqrt{n+1}$ (or $\Delta
\sqrt{n}$). Thus the \textquotedblleft independent LZ transition
approximation\textquotedblright\ holds as long as the Fock states $%
\left\vert n\right\rangle $, with $n>4\omega ^{2}/\Delta ^{2}$, are not
occupied, i.e., when $\omega >\max \{\sqrt{v}/4,\left\vert \alpha
\right\vert \Delta \}$.

The appearance of the second set of avoided-level crossings then allows the
occurrence of the second transition. Therefore, two stages of the LZ
transition are predictable and numerically shown in Fig.\thinspace
4\thinspace (b). We choose $\omega =10$ and $\Delta =0.5$ (in units of $%
\sqrt{v}$), and then there exist two (almost) independent LZ transitions.
The second transitions just occur in the vicinity of the second set of
avoided crossings at $t=2\omega /v$. The analytical results for $P_{\text{LZ}%
}(\infty )$ [in Eq.\thinspace (\ref{P_LZ_YS})] within the RWA are marked by
dashed lines. For\ different $\left\vert \alpha \right\vert ^{2}$ cases, the
first-stage LZ probabilities nicely agree with the RWA results. However,
after the second stage, the final LZ probabilities significantly deviate
from the RWA results.

In order to reveal the richness of the dynamics without the RWA, in
Fig.\thinspace 5, we have numerically calculated LZ probabilities for $%
\left\vert \alpha \right\vert ^{2}=1$ and various values of the couplings $%
\Delta $ and frequencies $\omega $. Obviously, once small frequencies such
as $\omega =1$ are chosen,\ the time $t_{\text{cross}}=2\omega /v$ is of the
order $\tau _{\text{LZ}}$, where the two LZ transitions start to interfere
with each other. By increasing the frequency $\omega $, there are two
visible stages of the LZ transition. The first-stage LZ probabilities do not
depend strongly on the frequency $\omega $, as long as $\left\vert \alpha
\right\vert \Delta \ll \omega $ is satisfied. However, the second-stage LZ
probability is influenced strongly by the frequencies $\omega $.

If all the avoided crossings are well separated, we can approximately treat
the transitions as being independent and compute the transition
probabilities as joint probabilities \cite%
{Ind-crossing1,Ind-crossing2,Ind-crossing3}. For the Hamiltonian (\ref%
{H-nonRWA}) without the RWA, the final probability to find the TLS at $%
\left\vert \uparrow \right\rangle $ from the initial state $\left\vert
\uparrow n\right\rangle $ is \cite{Ind-crossing3}
\begin{equation}
P_{\uparrow ,n\rightarrow \uparrow }=P_{\uparrow ,n-1}P_{\uparrow ,n}+\left(
1-P_{\uparrow ,n-1}\right) \left( 1-P_{\uparrow ,n-2}\right) ,
\end{equation}%
where $P_{\uparrow ,n}=\exp [-\pi \Delta ^{2}(n+1)/2v]=P_{\uparrow ,0}^{n+1}$%
, and the expanded form of the equation above still holds for the $n=0$ and $%
1$ cases. By substituting $P_{\uparrow ,n\rightarrow \uparrow }$ into $P_{%
\text{LZ}}\left( \infty \right) =1-\sum_{n=0}^{\infty }\left\vert
C_{n}\right\vert ^{2}P_{\uparrow ,n\rightarrow \uparrow }$, we obtain the LZ
probability in the independent-transition approximation and weak couplings:
\begin{equation}
P_{\text{LZ}}\left( \infty \right) =\frac{K_{\alpha ,\theta }}{P_{\uparrow
,0}}\left[ f_{\alpha ,\theta }(P_{\uparrow ,0})-f_{\alpha ,\theta
}(P_{\uparrow ,0}^{2})\right] ,  \label{P-nonRWA-CS}
\end{equation}%
where the coefficient $K_{\alpha ,\theta }=e^{-\left\vert \alpha \right\vert
^{2}}/(1+\cos \theta e^{-2|\alpha |^{2}})$ depends on $\left\vert \alpha
\right\vert ^{2}$ and $\theta $. The function $f_{\alpha ,\theta }$ is
defined by $f_{\alpha ,\theta }(x)=\left( 1+x\right) \left( e^{\left\vert
\alpha \right\vert ^{2}x}+\cos \theta e^{-\left\vert \alpha \right\vert
^{2}x}\right) $. When $\left\vert \alpha \right\vert ^{2}\rightarrow 0$, for
$\cos \theta \neq -1$,\ we find that\ $P_{\text{LZ}}\left( \infty \right) $
tends to the standard LZ probability $1-P_{\uparrow ,0}$, whereas, for $\cos
\theta =-1$, $P_{\text{LZ}}\left( \infty \right) $ approaches $1-P_{\uparrow
,0}^{3}$.\ In the large photon number limit $\left\vert \alpha \right\vert
^{2}\rightarrow \infty $, Eq.\thinspace (\ref{P-nonRWA-CS}) gives $P_{\text{%
LZ}}\rightarrow 0$.

\begin{figure}[tbp]
\includegraphics[width=0.5\textwidth,height=0.38\textwidth]{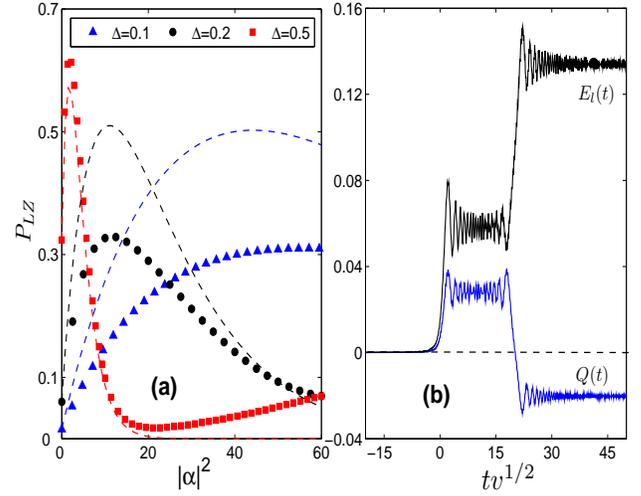}
\caption{(Color online) (a) Without the RWA, the long-time LZ probability $%
P_{\text{LZ}}$ as a function of $|\protect\alpha |^{2}$ for different
couplings $\Delta $, in units of $\protect\sqrt{v}$. The frequency $\protect%
\omega =10$ (in units of $\protect\sqrt{v}$) and the numerical time
integration is performed over $[-50,50]$. The blue solid triangles, black
solid circles and red solid squares denote the numerical results for $%
\protect\omega =0.1,~0.2$, and $0.5$, respectively. The dashed lines
correspond to the analytical results of Eq.~(\protect\ref{P-nonRWA-CS}), and
the blue, black and red lines denote $\protect\omega =0.1,~0.2$ and $0.5$,
respectively. (b) Linear entropy $E_{l}(t)$ and Mandel parameter $Q(t)$ as a
function of time (in units of $1/\protect\sqrt{v}$). The average photon
number is $|\protect\alpha |^{2}=1$, the coupling $\Delta =0.1$, and the
frequency $\protect\omega =10$, in units of $\protect\sqrt{v}$.}
\end{figure}
By performing a numerical time integration for $t\in \lbrack -50,50]$, in
units of $1/\sqrt{v}$, we calculate the long-time LZ probability as a
function of $\left\vert \alpha \right\vert ^{2}$ for different values of $%
\Delta $; these results are shown in Fig.\thinspace 6\thinspace (a).
Clearly, the LZ probability possesses a nonmonotonic behavior. When the
coupling is sufficiently weak, such as $\Delta =0.1$, the LZ probability
increases with $\left\vert \alpha \right\vert ^{2}$ monotonically in quite a
long region of $\left\vert \alpha \right\vert ^{2}$. However, a stronger
coupling like $\Delta =0.5$ makes the LZ probability achieve a maximum value
quickly. Unfortunately, the approximate result in Eq.\thinspace (\ref%
{P-nonRWA-CS}) is not in good agreement with the numerical results, which is
unlike the thermal state case\thinspace \cite{Ind-crossing3}. This is
because the initial photon state is of Poissonian (sub- or super-)
statistics, and the independent-transition condition $\omega >\max \{\sqrt{v}%
/4,\left\vert \alpha \right\vert \Delta \}$, will be destroyed by increasing
$\left\vert \alpha \right\vert $. Moreover, the LZ process strongly depends
on the frequencies $\omega $ even in the weak-coupling region (see
Fig.\thinspace 5). Nevertheless, the approximation result of Eq.\thinspace (%
\ref{P-nonRWA-CS}) still indicates a significant fact that the long-time LZ
probability no longer increases monotonously with the photon number $%
\left\vert \alpha \right\vert ^{2}$.

In Fig.\thinspace 6\thinspace (b), the dynamics of entanglement and Mandel
parameter $Q$ can also present two transitions. From all the numerical
results shown in Figs.~4--6, we find that there are some qualitative
differences between the results within and without the RWA. Nevertheless,
under certain conditions, the RWA results are in good agreement with the
first-stage LZ transition. Thus one can choose properly weak couplings $%
\Delta $ and large frequencies $\omega $ to extend the time interval between
the two LZ transitions, when the RWA is valid.

For comparison, we also consider the thermal state case, and below we
calculate the LZ probability with the RWA and without the RWA.

\subsection{Thermal state case}

When the photon field initially starts from a thermal state, the density
matrix of the total system is
\begin{equation}
\rho _{\text{tot}}\left( 0\right) =\left \vert \uparrow \right \rangle \left
\langle \uparrow \right \vert \otimes \frac{1}{Z}\exp (-\omega a^{\dagger
}a/T),
\end{equation}%
where the partition function$\ Z=(1-e^{-\omega /T})^{-1}$ (setting $\hbar
,k_{B}=1$). Within the RWA, the final LZ transition probability becomes
\begin{figure}[tbp]
\includegraphics[bb=23 345 555 589, width=0.44\textwidth]{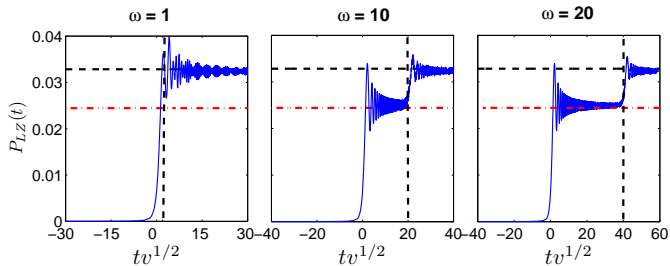}
\caption{(Color online) The initial state of the photon field is now a
thermal state. Without the RWA, we plot the LZ probability $P_{\text{LZ}}(t)$
as a function of time in units of $1/\protect\sqrt{v}$ for various
frequencies $\protect\omega =1,~10,$ and $20$, in units of $\protect\sqrt{v}$%
. The coupling $\Delta =0.1$ in units of $\protect\sqrt{v}$, and the scaled
temperature $T/\protect\omega =1$. The horizontal red dash-dot lines mark
the analytical results in Eq.~(\protect\ref{P-RWA-Thermal}) with the RWA,
and the horizontal black dashed line corresponds to the analytical results
in Eq.~(\protect\ref{P-nonRWA-Thermal}) without the RWA. The vertical dashed
lines indicate the time between the successive two stages of the LZ
transitions.}
\end{figure}
\begin{equation}
P_{\text{LZ}}\left( \infty \right) =\frac{1-P_{0,\text{LZ}}}{1+\bar{n}P_{0,%
\text{LZ}}},  \label{P-RWA-Thermal}
\end{equation}%
where the average photon number $\bar{n}=[\exp (\omega \beta )-1]^{-1}$, and
$P_{0,\text{LZ}}$ denotes the standard LZ probability: $P_{0,\text{LZ}%
}=1-\exp (-\pi \Delta ^{2}/2v)$. When the scaled temperature $T/\omega
\rightarrow 0$, we find that $P_{\text{LZ}}\left( \infty \right) $ tends to
the standard LZ probability.

\textit{Without} the RWA, the\ approximate final LZ transition probability
for finite temperature and weak coupling becomes\thinspace \cite%
{Ind-crossing3}
\begin{equation}
P_{\text{LZ}}\left( \infty \right) =\frac{G_{T}}{P_{\uparrow ,0}}\left[
f_{T}(P_{\uparrow ,0})-f_{T}(P_{\uparrow ,0}^{2})\right] ,
\label{P-nonRWA-Thermal}
\end{equation}%
where $G_{T}=1-\exp (-\omega /T)$, and the function $f_{T}(x)=\left[ 1-x\exp
(-\omega /T)\right] ^{-1}$. For the initial thermal state, the probability $%
p(n)$ of finding $n$ photons is exponentially dependent on $\omega $, which
is quite different from the case of the YS coherent state, where the photon
distribution is Poissonian and independent of $\omega $. Consequently, in
Fig.\thinspace 7, for an initial thermal state, we plot the LZ probability
versus time for weak coupling $\Delta $. The first-stage LZ probability is
consistent with the RWA results, and the second-stage LZ transitions also
confirm the approximate results of Eq.\thinspace (\ref{P-nonRWA-Thermal}).
Moreover, for different frequencies $\omega $, although the curves differ
strongly around $t=0$, the LZ probabilities converge toward the same value,
which is significantly different from the case of the\ YS coherent state,
where the LZ probabilities strongly depend on $\omega $ and do not converge
to a single value (see Fig.\thinspace 5).

\section{Conclusions}

We have investigated the LZ transition in a composite system of a TLS
coupled to a single-mode photon field. The initial state of the field was
chosen as a superposition of coherent states. Within the RWA, we
analytically obtained the LZ probability as a function of the average photon
number and the superposition parameter. By increasing the average photon
number and choosing proper superposition parameters, one can enhance the LZ
probability. We also found that both the creation entanglement (between the
TLS and field) and the photon distribution change drastically when the LZ
transitions occur, which is helpful for revealing the effects of the LZ
transition on the whole quantum system.

Beyond the RWA, we found some qualitative differences from the RWA results.
The final LZ probability no longer monotonically depends on the average
photon number. In addition, two obvious stages of the LZ transition appear
in the vicinity of the successive avoided crossings, and the RWA results can
only indicate the first-stage LZ probability. The final LZ probability,
after the second stage, significantly deviates from the RWA results and
strongly depends on the frequencies $\omega $, even for weak couplings $%
\Delta $.

We found that the LZ dynamics is quite different from the thermal state
case, which is due to the Poissonian distribution in the superposition of
coherent states. Although the RWA fails in estimating the final LZ
probability, it plays an important role in characterizing the finite-time
coherent oscillations generated by the LZ transitions. With sufficiently
weak coupling $\Delta $ and large frequency $\omega $, one can extend the
time interval between the two LZ transitions when the RWA is accurate
enough. Very recently, the authors of Ref.\thinspace \cite{Larson} found the
absence of vacuum-induced Berry phases without the RWA. Our results also
provide examples indicating that the RWA leads to faulty results.

\acknowledgments We would like to thank S. N. Shevchenko for useful
discussions. X.G.W. acknowledges support from the NFRPC under Grant
No.~2012CB921602 and from the NSFC under Grant No.~11025527 and No.
10935010. F.N. is partially supported by the ARO, NSF Grant No. 0726909,
JSPS-RFBR Contract No. 12-02-92100, Grant-in-Aid for Scientific Research
(S), MEXT Kakenhi on Quantum Cybernetics, and the JSPS via its FIRST
program. Z.S. acknowledges support from the NSFC under Grant No.~11005027,
the Natural Science Foundation of Zhejiang Province with Grant No.\thinspace
Y6090058, and Program for HNUEYT under Grant No.\thinspace HNUEYT
2011-01-011. J.M. acknowledges support from the Scholarship Award for
Excellent Doctoral Student granted by the Ministry of Education.

\end{document}